# Investigating polaron formation in anatase and brookite TiO$_2$ by density functional theory with hybrid-functional and DFT+U methods

Jeffrey Roshan De Lile,[1] Sung Gu Kang,[2] Young-A Son[3] and Seung Geol Lee[1,*]

[1] *Department of organic material science and engineering, Pusan National University 2, Busandaehak-ro 63beon-gil, Geumjeong-gu, Busan 46241, Republic of Korea*

[2] *School of Chemical Engineering, University of Ulsan, 93 Daehak-ro Nam-gu, Ulsan 44610, Republic of Korea*

[3] *Department of Advanced Organic Materials Engineering, Chungnam National University, 220 Gung-dong, Yuseong-gu, Daejeon 305-764, Republic of Korea*

---

*Corresponding author:
seunggeol.lee@pusan.ac.kr (S.G. Lee)

**Abstract**

Anatase and brookite are robust materials with enhanced photocatalytic properties. In this study, we used density functional theory (DFT) with a hybrid functional and the Hubbard on-site potential method to determine electron- and hole-polaron geometries for anatase and brookite, and their energetics. Localized electron and hole polarons were predicted not to form in anatase using DFT with hybrid functionals. In contrast, brookite formed both electron and hole polarons. The brookite electron-polaronic solution exhibits coexisting localized and delocalized states, with hole polarons mainly dispersed on two-coordinated oxygen ions. Hubbard on-site potential testing over the wide 4.0–10 eV range revealed that brookite polarons are formed at U = 6 eV, while anatase polarons are formed at U = 8 eV. The brookite electron polaron was always localized on a single titanium ion under the Hubbard model, whereas the hole polaron was dispersed over four oxygen atoms, consistent with the hybrid DFT studies. The anatase electron polarons were dispersed at lower on-site potentials but were more localized at higher potentials. Both methods predict that brookite has a higher driving force for the formation of polarons than anatase.

# 1. Introduction

Titania (TiO$_2$) is used in a variety of applications, including sensors,[1-3] photocatalysis,[4-7] dye-sensitized solar cells,[8-10] and has been the popular choice of material in the aforementioned applications due to its low cost, lack of toxicity, and thermal stability. Common TiO$_2$ polymorphs, namely rutile, anatase, and brookite, exhibit different photocatalytic activities, irrespective of their identical chemical compositions. The most stable rutile and anatase polymorphs have been extensively investigated, both experimentally[11-14] and theoretically,[15-17] in order to understand their differences in photocatalytic activity. Due to the well-known difficulties associated with the synthesis of pure brookite titania, the literature related to the brookite polymorph is limited.[18] A recent study pointed-out that electron and hole trapping in titania polymorphs play vital roles in photocatalytic activity.[19] The authors emphasized that trapping depth was crucial for photocatalysis performance, and reported that brookite has a moderate trapping depth; hence brookite is active in photocatalytic reductions and oxidations. In contrast, anatase has a very shallow trapping depth, which reduces the lifetimes of photo-generated electrons and holes due to recombination.[19] Therefore, elucidating the mechanism of electron and hole formation in titania polymorphs is very important for developing an understanding of the catalytic properties of these materials.

The formation of photogenerated electrons and holes is accompanied by lattice vibrations (phonons) that are quasiparticles collectively known as "polarons".[20] Therefore, polaron formation is always associated with lattice distortion. The pioneering theoretical work of Deskins et al.,[21-23] provided detailed information of the structures of electron and hole polarons in both rutile and anatase polymorphs based on the Hubbard on-site potential method. Nevertheless, they acknowledged the fact that no universal Hubbard U parameter is capable of reproducing all of the experimental observations; depending on the properties of interest, the most suitable U value for the model must be obtained by trial and error. In that work, the authors used 10 eV as the effective Hubbard parameter for investigating the electron- and hole-transport properties in both rutile and anatase. However, recent studies using the constrained-random-phase approximation proposed effective U values of 3.9 eV for rutile and 4.1 eV for anatase.[24] These authors found that the excess electron was homogeneously distributed (delocalized); consequently electron-polaron formation is unfavorable in a perfect anatase crystal lattice; moreover, they claimed that anatase forms electron polarons next to oxygen vacancies on its surface. Selloni et al.[25] intensively studied electron- and hole-polaron formation in the anatase polymorph using the B3LYP hybrid functional. They also proposed electron-polaron formation on the anatase surface due to the higher trapping energy at the (101) surface than in bulk anatase.

Mo and Ching[26] reported the electronic structure and optical properties of the brookite polymorph (as well as rutile and anatase) using the orthogonalized linear-combination-of-atomic-orbitals method in the local density approximation (LDA). These workers were aware of the electron self-interaction error and introduced a self-interaction correction and Green's-function-based quasiparticle terms to correct this issue. Therefore, screened Coulomb potentials that include range-separated hybrid DFT functionals appear in the literature for studying titania phases. McKenna et al.[27] recently reported charge-transfer levels (CTLs) for several polymorphs of titania, including rutile, anatase, brookite, $TiO_2$ (H), $TiO_2$ (R), and $TiO_2$ (B) using their own hybrid-functional based on the generalized Koopmans condition to minimize the self-interaction error of electrons. The authors showed hole-polaron formation in anatase and brookite, however electron-polaron formation was not observed. Consequently, the different photocatalytic activities are naturally attributed to the formation of different polarons in the titania phases. However, electron paramagnetic resonance results showed that both anatase and brookite phases contain reduced $Ti^{3+}$ centers, which are precursors for electron-polaron formation.[28] Van de Vondele et al.[29] reported that electron-polaron formation in titania depends on the system size and the amount of exact exchange. Hence, they proposed the application of the random phase approximation (RPA) on top of PBE0 hybrid orbitals with a 30% exact exchange. In addition, an 864-atom supercell was suggested as the best compromise between computational time and accuracy for calculating the electron-polaron formation energy of anatase; these are highly computationally demanding or prohibitively expensive tasks using the current infrastructure of most computing centers.

The recent literature[19, 30-32] provides experimental evidence for the higher photocatalytic activities of the pure and mixed phases of brookite. However, a lack of fundamental understanding about the polaron structure of the brookite phase hinders the development of visible-light-active photocatalyst based on brookite. According to our knowledge, no account of the polaron structure of the brookite phase of titania has been reported to date. In this study, we investigated the electron- and hole-polaron structures of the bulk phases of anatase and brookite using the PBE0 hybrid functional and Hubbard on-site potential method. In order to reduce computational costs, we deliberately used a 48-atom system for comparing the anatase and brookite results. Based on our findings, we explain the experimental trends currently observed for the anatase and brookite systems.

## 2. Computational Details

The electronic structures of pure anatase and brookite stoichiometric bulk phases were studied using conventional cells and 48-atom supercells using the plane-wave DFT code in the Vienna *ab initio* software package (VASP).[33-34] Core electrons were described by the projector-augmented-wave (PAW) method with a 650 eV cutoff.[35] Here, we used large cores for both the Ti and O atoms, with valence electron configurations of $3d^24s^2$ and $2s^22p^4$, respectively. Our large core approximation is justified by the calculated band gap values that agreed well with the optical band gap of these materials. Initially, orbitals were calculated with the PBE-GGA exchange-correlation functional[36-37] with tight criteria for the allowed errors in total orbital energies (EDIFF = $10^{-8}$), and a 5 × 5 × 5 Monkhorst-pack grid.[38] Frequency calculations were subsequently performed to check the convergence and to confirm the sufficiency of above-mentioned Monkhorst-pack grid for these systems. These DFT orbitals were used as the starting points for PBE0-hybrid calculations.[39] Exact exchanges of 25%, 30%, and 35%, with a 100% PBE correlation energy were used in the PBE0-hybrid calculations. A more-robust normal algorithm with linear mixing and Kerker mixing parameters[40] was employed for calculations with the PBE0 hybrid functional. Moreover, the "number of valence electrons" tag was used to define the number of electrons in spin-polarized calculations of polaron-formation energies. One excess electron was added when studying electron polarons, and one electron was removed from the system when investigating hole polarons.

Suitable Hubbard parameters were examined in the 4–10 eV range for both electron and hole polarons. We used Dudarev's approach as implemented in the VASP.[41] Although on-site Coulomb (U) and exchange (J) parameter differences (U-J) are meaningful under this scheme, the VASP implementation effectively sets the J value to zero. Therefore, the selected U value can be considered to be the effective U ($U_{eff}$). We applied on-site Hubbard U correction to titanium 3d orbitals and oxygen 2p orbitals to simulate electron polarons and hole polarons, respectively. According to the literature[27] and our PBE0 hybrid calculations, both anatase and brookite have a tendency to efficiently produce hole polarons. Thus, hole polarons have higher polaron formation/trapping energies than those of electron polarons in these materials. Higher hole-polaron energies compared to electron-polaron energies were only observed for a Hubbard U value of 8 eV for both polymorphs. Hence, a $U_{eff}$ of 8 eV was chosen to compare energies using the DFT + U method. However, we would like to emphasize that the U values presented here may valid only for the 48 atom supercells in the VASP code simulation as the Hubbard potential severely depends on the DFT code.[42]

Three basic calculations were performed for all defect-free supercells. First, each supercell was optimized with the PBE-GGA functional. Next, the calculation was switched to accommodate the PBE0 range-separated hybrid functional and re-optimized from the PBE-GGA-functional parent orbitals. However, we did not optimize the cell parameters at this step and only ionic relaxation was computed. During the third step, an electron (or a hole) was added to the supercell. Here, we did not additionally attempt to localize the electron (or the hole) by introducing defects around the respective atom to form a precursor potential. Thus, all the localized polarnic states spontaneously occurred during optimization. In this step, we first computed the delocalized polaron on the non-relaxed geometry ($E_{deloc,non-relax}$), after which those structures were fully optimized to obtain localized polarons in the relaxed geometry ($E_{loc,relax}$). We performed the same number of steps in the DFT+U method with the respective $U_{eff}$ value. Hence, the polaron-formation/trapping energy was calculated using the following equation[24, 29, 43].

$$E_{pol} = E_{deloc,non-relax} - E_{loc,relax} \qquad (1)$$

## 3. Results and Discussion

3.1 Equilibrium structures

The model structures used in this study are displayed in Figure 1, and all of the bond lengths and lattice constants are listed in Table 1 for unperturbed supercells that were investigated with the PBE (GGA level of theory) and hybrid PBE0 functionals. Hence, our results can be compared with available experimental data to validate our models. Both the PBE- and PBE0-

functional-calculated structural parameters and interatomic Ti-O bond distances of brookite and anatase are in good agreement to those obtained experimentally.

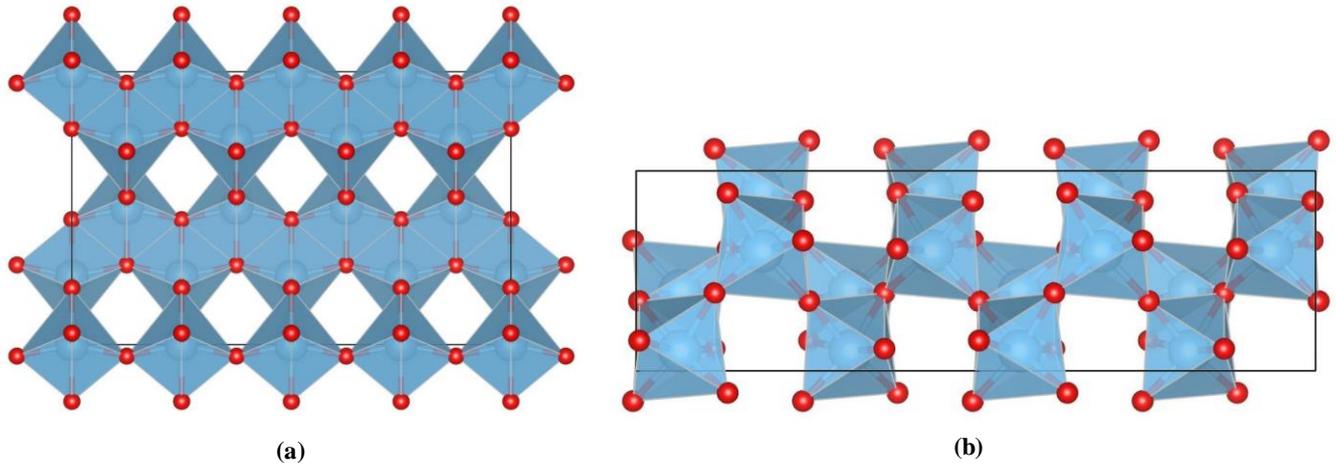

Figure 1. Model structures with 48 atoms of (a) anatase and (b) brookite. Blue and red represent titanium and oxygen, respectively. Anatase: tetragonal, space group I41/amd–$D_{4h}^{19}$, a = b = 3.776 Å and c = 9.486 Å; brookite: orthorhombic space group Pbca-$D_{2h}^{15}$, a = 9.166 Å, b = 5.436 Å and c = 5.135 Å. Anatase viewed in the xz plane, and brookite viewed in the xy plane.

Table 1. Cell parameters and interatomic distances in anatase and brookite calculated using the PBE and PBE0 functionals.[a]

| | Anatase | | | |
|---|---|---|---|---|
| | a/Å | b/Å | c/Å | Ti-O/Å |
| Experimental[44] | 3.782 | 3.782 | 9.502 | 1.932;1.978 |
| PBE | 3.776 | 3.776 | 9.486 | 1.927;1.988 |
| PBE0 | 3.776 | 3.776 | 9.486 | 1.930;1.974 |

| | Brookite | | | |
|---|---|---|---|---|
| | a/Å | b/Å | c/Å | Ti-O/Å |
| Experimental[44] | 9.182 | 5.456 | 5.143 | 1.923;1.930;1.990;1.999 1.863;2.052 |
| PBE | 9.166 | 5.436 | 5.135 | 1.922;1.935;1.969;2.005 1.881;2.032 |
| PBE0 | 9.166 | 5.436 | 5.135 | 1.925;1.932;1.975;1.980 1.874;2.032 |

[a]Both equatorial and axial interatomic (Ti-O) bond lengths are listed. For, brookite all four equatorial bond lengths are listed first, with the two axial bond lengths listed below them. Note, cell parameters from DFT+U calculations are not reported here.

## 3.2 Bandgaps

Initial attempts to calculate the bandgap of brookite using theoretical techniques were reported by Grätzel and co-workers,[45] who used extended Hückel molecular orbital theory. They calculated bandgaps of 3.14, 3.23, and 3.02 eV for brookite, anatase, and rutile, respectively, which revealed that brookite has a gap value that lies between those of the other two polymorphs of titania. Bandgaps were recently calculated using PBE, range-separated hybrid

(HSE06), and Green's-function-based quasiparticle ($G_0W_0$) approaches in the VASP code. The predicted energies of 1.86 eV (PBE), 3.30 eV (HSE06), and 3.45 eV ($G_0W_0$) for brookite are in very good agreement with those of an earlier study;[46] that study reported values for anatase of 1.94 eV (PBE), 3.60 eV (HSE06), and 3.73 eV ($G_0W_0$). Therefore, brookite has a lower bandgap than anatase. Experimentally determined bandgaps lie between 3.1 and 3.4 eV for brookite,[31] and the optical bandgaps of brookite and anatase are 3.3 eV[47] and 3.4 eV,[48] respectively.

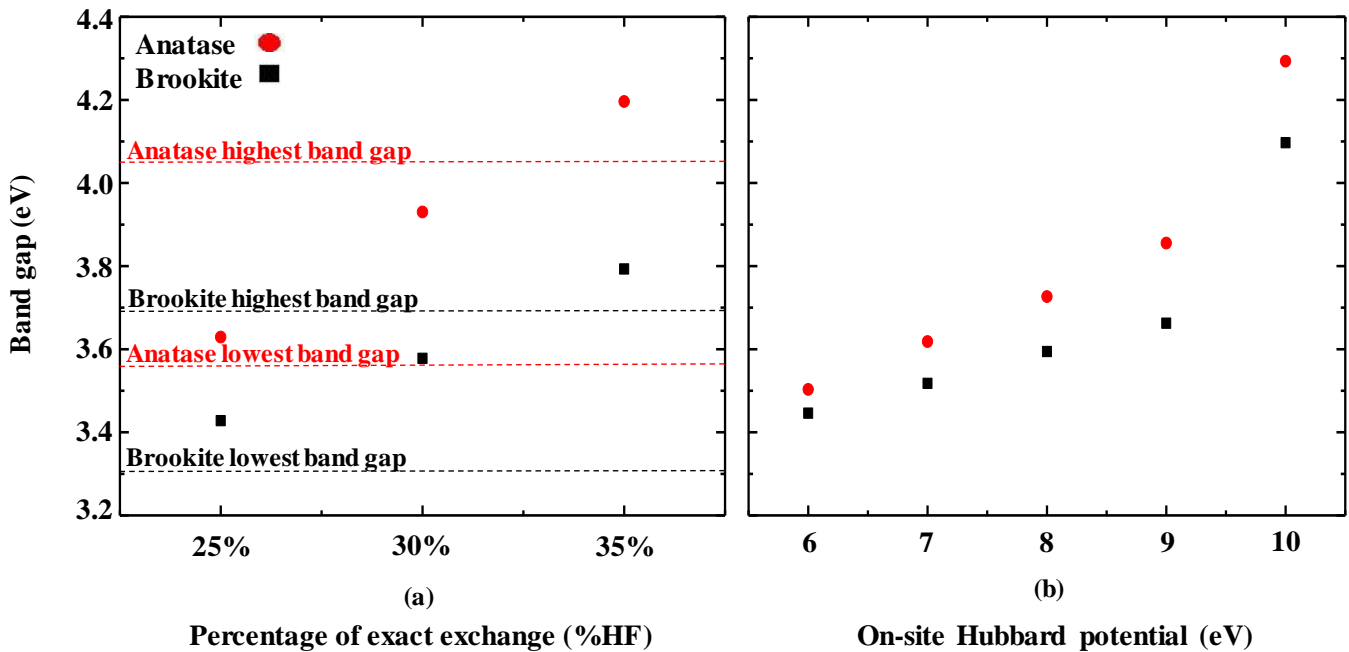

Figure 2. Bandgaps predicted by the (a) PBE0 and (b) on-site Hubbard models. Black dashed lines in (a) represent the literature-reported bandgaps (from ref [49]) calculated using hybrid or GW methods for brookite, while the red dashed lines represent literature-reported bandgaps for anatase. Anatase has higher bandgap than brookite, in agreement with literature reports and the experimentally determined optical bandgaps.

In this work, the bandgap of brookite was determined to be lower than that of anatase calculated using the PBE0 hybrid functional and the on-site Hubbard potential (Figure 2). These results are in good agreement with the above-mentioned literature[49] and lend credibility to this study. The hybrid functional with 25% exact exchange and an on-site potential value of 6 eV provided a bandgap of ~3.4 eV for brookite, which is in good agreement with the experimental optical gap. The bandgap energy of anatase is clearly higher in each case, and the energy value was observed to increase linearly with increasing percentage of exact exchange and on-site Hubbard potential. However, the difference between the bandgap of brookite and anatase increased with increasing Hartree-Fock percentage (%HF); these differences are calculated to be 0.20, 0.35, and 0.37 eV for exact exchanges of 25%, 30%, and 35%, respectively, which we attribute to the small lattice sizes of the titania polymorphs that may be associated with quantum-confinement effects. Nevertheless, the differences between the brookite and anatase bandgaps are calculated to be minimal using the on-site Hubbard potentials; they range from 0.06 eV (U = 6 eV) to 0.19 eV (U = 10 eV). Except for a potential of 10 eV, these calculated bandgaps also increase linearly with increasing U value, which is attributed to higher bandgap opening due to the compellingly higher on-site potential. The red and black dashed lines in Figure 2 indicate the literature reported[49] bandgaps from higher-level calculations for anatase and brookite, respectively. The bandgaps are within the expected range for large supercell calculations.

3.3 Polaron geometries by hybrid DFT

It is well-known fact that hybrid functionals can produce both localized and delocalized solutions for excess-electron and electron-deficient structures.[29] A percentage of exact exchange (Hartree-Fock, HF) is applied to the simulation to stabilize either the localized or delocalized state. Generally, a lower percentage of exact exchange favors a delocalized solution and *vice versa*. This is mainly due to the metastable nature of the localized solution to the polaron structure at low %HF; hence structure optimization leads to a delocalized solution. Nevertheless, the delocalized solution becomes increasingly metastable with higher amounts of exact exchange such that a localized solution is eventually found at the end of the simulation. We examined anatase clusters with 25%, 30%, and 35% HF; however, polarons (both electron and hole) were not formed in the bulk unit cells (the bulk anatase structures with fully delocalized electron and hole polarons can be found in Figure S1 of the Supporting Information). This is attributed to the system size, as it is commonly reported that polarons in anatase are large or Fröhlich polarons that are spread across several lattice units. Therefore, at the higher concentration limit, excess trapped electrons are metastable, which always leads to the delocalized solution. This behavior is in agreement with experimental observations reported in the literature.[50] In addition, Moser et al.[50] reported that photon adsorption by ground-state anatase produces a large hole polaron with a radius of 20 Å, which is dissociated throughout the lattice at higher electron density. Moreover, recent theoretical work reported that polaron formation is unfavorable in a stoichiometrically perfect anatase lattice, but is favored in rutile.[24] Another study reported that no localized solution exists for an excess-electron anatase simulation even with an exact exchange of 30%.[29] These authors reported that stable polaron structures only were observed in anatase with large unit cells, and lattice relaxation was observed to spread over more than four unit cells (~15.2 Å). As opposed to anatase, brookite stoichiometric lattices with 48 atoms simulated in this work with the hybrid PBE0 functional produced both electron and hole polarons.

3.3.1 Electron polarons in brookite

The excess electron in brookite tends to localize itself at two separate Ti ions in the unit cell. The bond lengths are slightly affected by the percentage of exact exchange employed; thus, they are weakly dependent on the amount of exact exchange. We compared the results obtaining using the lowest (25% HF) and highest (35% HF) configurations to highlight the above observation. The average Ti-O bond lengths and the changes resulting from polaron formation are reported in Tables S1 and S2 of the Supporting Information for both the hybrid and on-site Hubbard models, respectively. Figure 3 displays the electron-polaron geometries obtained from the hybrid calculations. Axial or apical bonds refer to the bonds linking the two positions that are collinear with the central Ti ion, while equatorial bonds refer to the bonds passing through the central Ti ion and are orthogonal to the axial bonds. We use the above definitions to describe and differentiate the bond-length variations observed during polaron formation in this work. Lattice relaxation was observed throughout the brookite in the [100] and [010] directions of the supercell (~18.33 Å). However, the longest bond from Ti to an equatorial oxygen (Ti-$O_{eq}$), was lengthened by +0.06 Å (and +0.07 Å) with respect to that of the unperturbed structure at 25% (and 35%) exact exchange. The shortest Ti-$O_{eq}$ bond contracted by -0.04 Å and -0.05 Å at 25% and 35% exchange, respectively, whereas the longest bond from Ti to an axial oxygen (Ti-$O_a$) was minimally perturbed and the shortest Ti–$O_a$ bond was elongated by +0.02 Å and +0.03 Å at 25% and 35% exact exchange, respectively.

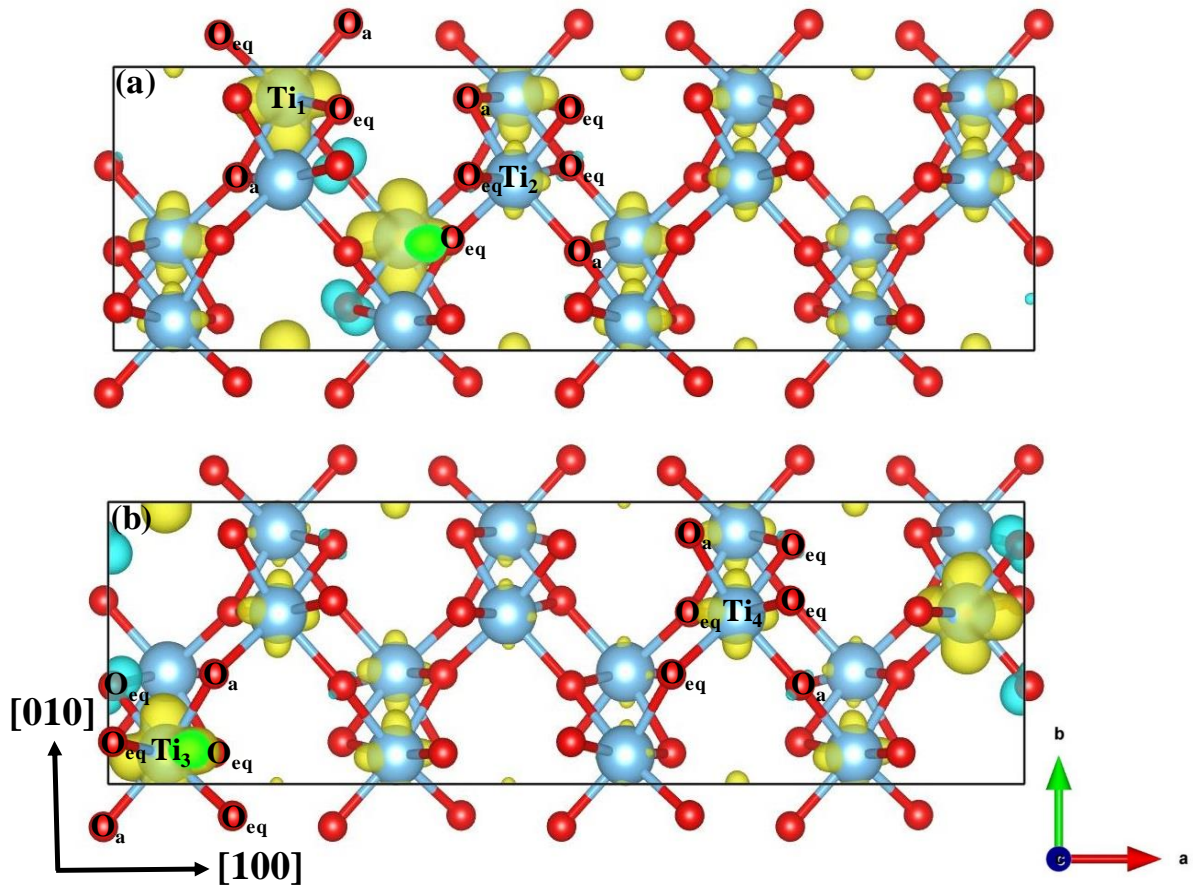

Figure 3. Hybrid-functional calculated brookite supercells with electron polarons calculated at (a) 25% and (b) 35% exact exchange. Note that the electron polaron is localized on two Ti ions and the electron-polaron wavefunction is delocalized over other Ti ions. Color scheme: blue, titanium; red, oxygen; yellow, polaronic wavefunction at an isosurface value of 0.003 e/Bohr$^3$. Ti-O bond distances are reported for equatorial ($O_{eq}$) and axial ($O_a$) oxygen bonds at the polaronic site and away from it (see Table S1).

Ti-O bond-length variations measured away from the localized polarons ranged between +0.02 Å and -0.02 Å; hence, these bond lengths are only slightly different to those of the unperturbed geometry. We conclude, therefore, that no fully localized electron polaron is formed at other Ti ions. On the basis of the more-delocalized nature of the electron-polaron wavefunction in the brookite lattice (except in the two fully localized states), we believe that both localized and delocalized states coexist in this lattice. However, we observed a delocalized polaronic solution even at the highest exact exchange used (35%) in this study. Therefore, a more-stable localized solution must be obtainable at higher exact exchanges (40% or above); however, the electron polaron can also be artificially localized at higher exact exchanges,[29] therefore we did not attempt to apply a higher %HF. Nevertheless, the polaron-formation energies were determined to be negative; therefore electron polarons are even stable in these supercells. Furthermore, the Ti-O bond lengths increase in the direction of the a-lattice constant (9.166 Å), while they decrease in the direction of the c-lattice constant (5.135 Å), which provides a certain level of anisotropy to the coordination environment at the charge-localization center.

3.3.2 Hole polarons in brookite

Looking at the polaronic wavefunction, it is clear that four hole polarons are localized at four oxygen atoms in brookite (Figure 4 (a) and (b)). They are more pronounced, and three localized wavefunctions are observed at two-coordinated oxygen sites, with one localized at a three-coordinated oxygen site. Unlike for the electron polaron, the longest axial Ti-O bond length decreased by -0.03 Å and -0.04 Å using 25% and 35% exact exchange, respectively, with respect to the unperturbed structure. The shortest Ti-$O_a$ bond increased by +0.04 Å and +0.05 Å respectively, using 25% and 35% exact exchange. The equatorial bonds are distorted by between +0.01 Å and -0.02 Å; hence, the equatorial Ti-O bonds are less affected in the hole-polaron structure. The Ti-O bond lengths increase asymmetrically at the hole-polaron site. At the two-coordinated oxygen sites, variations between +0.01 Å and +0.05 Å were observed, while variations between +0.01 Å and +0.06 Å were observed at the three-coordinated oxygen site.

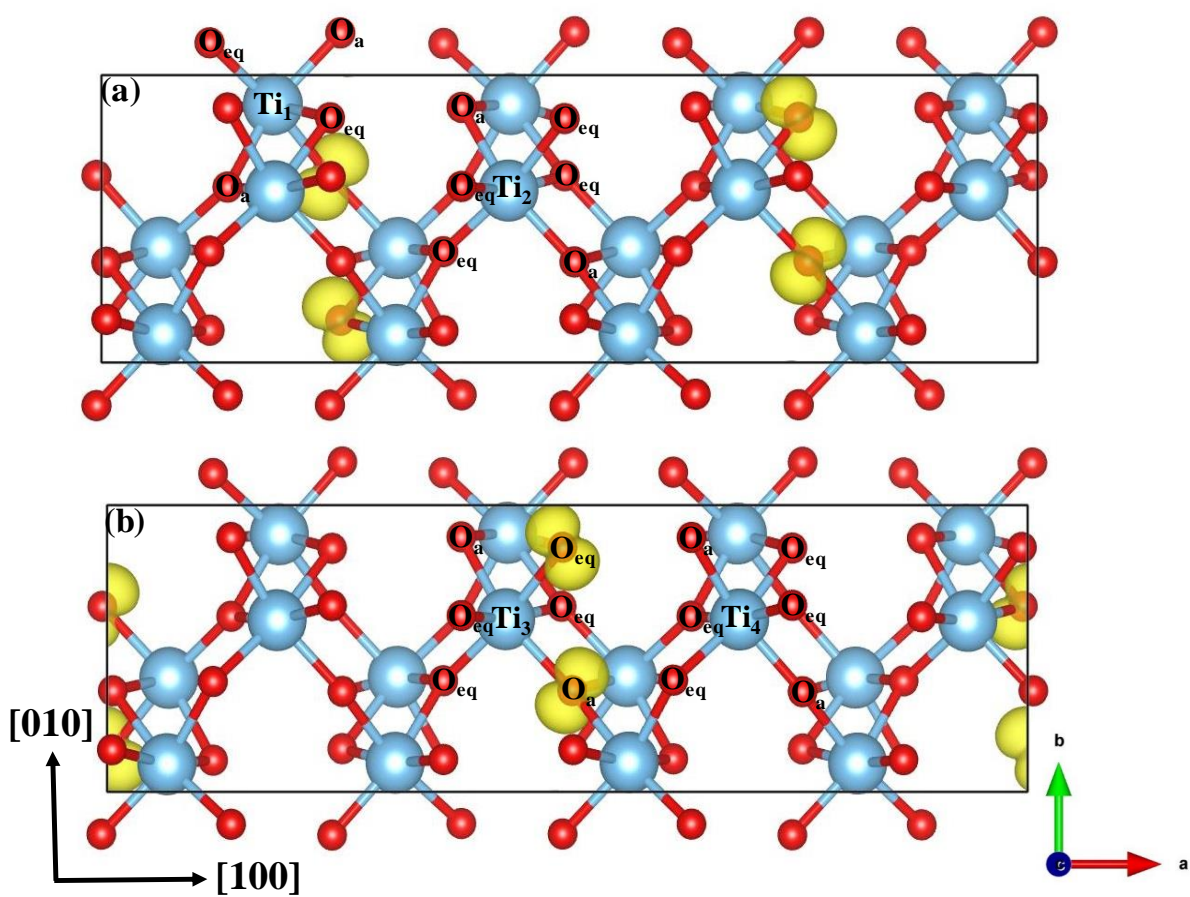

Figure 4. Brookite-supercell hole-polaron structures calculated using the hybrid functional with (a) 25% and (b) 35% exact exchange. Holes are localized on four oxygens, mainly the two-coordinated and three-coordinated oxygen ions. Color scheme: blue, titanium; red, oxygen; yellow polaronic wavefunction at an isosurface value of 0.003 e/Bohr$^3$. Ti-O bond distances are provided for equatorial and axial oxygen bonds at the polaronic site and away from it (see Table S1).

The hole-polaronic wavefunctions are fully localized on the oxygen ions. Therefore, brookite is able to produce a fully localized hole polaron in its bulk structure, which is in agreement with the recent work of McKenna and co-workers[27] who reported hole-polaron formation in brookite; however, no electron polarons were observed. The lack of electron polarons can be attributed to the coexistence of localized and delocalized solutions, as observed in this study. The brookite electron- and hole-polaron structures exhibit similar-scale lattice distortions that are spread across the entire lattice. Moreover, the electron-polaron geometry of brookite exhibits equatorial Ti-O bonds that are significantly distorted, while the two axial Ti-O bonds are significantly affected in the hole-polaron geometry; three bonds increase in length and three bonds decrease in length in the direction of the a-lattice vector (a = 9.166 Å), which induce anisotropy in the [100] direction.

3.4 Polaron geometries by DFT+U

The on-site Hubbard model is popular in the literature due to its low computational cost and convenience for tuning the properties of interest. However, it has an inherent drawback, namely that all properties of interest cannot be simulated with a single Hubbard on-site potential value.[21-22] In addition, it depends heavily on the implementation in DFT code.[42] Thus, we cannot directly compare the absolute values of U, which are taken from different DFT codes. Nevertheless, it can be used as a reasonable model for qualitatively comparing results with those from available theoretical or experimental studies. From the literature reported values of the on-site Hubbard parameter, U values from 4.0 eV[24] (VASP) to 10 eV[22-23] (VASP, CP2K) have been tested for anatase and brookite. Both electron and hole polarons are observed for brookite at U = 6 eV; however, both electron and hole polarons are observed for anatase at U = 8 eV. Nonetheless, an anatase hole polaron has also been observed at U = 7 eV. Therefore, bulk brookite may arguably form polarons rather more easily than bulk anatase. The brookite hole-polaron geometry was better reproduced at U = 6 eV when compared to that obtained from the PBE0 study, while the electron polaron was always fully localized on a single Ti atom, in contrast to that observed in the hybrid calculations.

3.4.1 Electron polarons in brookite

Figure 5 shows both electron- and hole-polaron geometries obtained from Hubbard on-site potential calculations. The electron polaron is fully localized at a single Ti atom in brookite. The four equatorial Ti-O bonds and the two axial Ti-O bonds are elongated from the Ti center.

The axial Ti-O bonds lengthen by about +0.16 Å (for the shortest bond) and +0.02 Å (for the longest bond). The longest axial bond seems to be resistant to change, and is minimally distorted, as was observed using the hybrid functionals. The equatorial Ti-O bonds lengthen by between +0.13 Å and +0.02 Å; therefore, all six bonds that form Ti octahedra dramatically change upon polaron formation. The structure stretches in the a- and c-lattice-vector directions, while the b-lattice-vector direction maintains the minimum distance. Hence, both the hybrid-functional and Hubbard models predict anisotropy in the electron polarons. Lattice relaxations are observed throughout the unit cell, as was observed in the hybrid functional study.

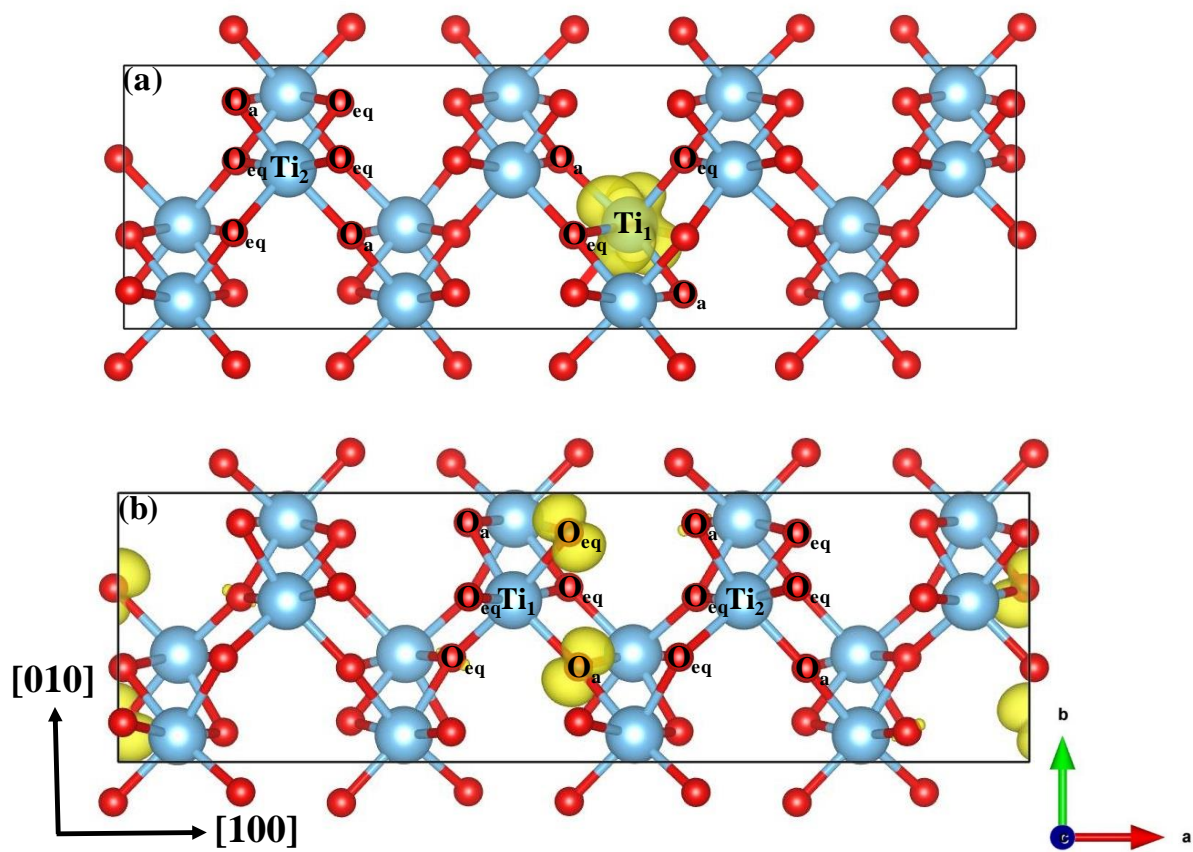

Figure 5. Hubbard-calculated brookite supercells with U = 6.0 eV: (a) electron and (b) hole-polaron structures. Note that the electron polaron is localized on a single Ti ion, whereas holes are localized on four oxygen ions. Color scheme: blue, titanium; red, oxygen; yellow, polaronic wavefunction at an isosurface value of 0.003 e/Bohr$^3$. Ti-O bond distances are recorded for equatorial and axial oxygen bonds at the polaronic site and away from it (see Table S2).

3.4.2 Hole polarons in brookite

The brookite hole-polaron geometry predicted by DFT + U is remarkably similar to that calculated with the PBE0 hybrid functional at 35% exact exchange. The hole-polaron structure is comparatively minimally distorted from the unperturbed structure. One of the axial Ti-O bonds with the shortest length was observed to increases (+0.04 Å) while the longest Ti-O$_a$ bond was observed decrease (-0.02 Å). Between the equatorial Ti-O$_{eq}$ bonds, two bonds increase in length (+0.03 Å, +0.01 Å) and two decrease in length (-0.02 Å), which clearly induces anisotropy in the a-lattice-vector direction, and provides a hole-migration path in the [100] direction. This reflects the recent experimental brookite literature, in which authors have reported the [210], [201], and [101] facets to be highly active surfaces for photocatalytic activity.[30] Thus, both the hybrid and Hubbard models predicted anisotropy in the hole-polaron geometry in the a-lattice-vector direction.

3.4.3 Electron polarons in anatase

The electron polarons in anatase are localized on three of the Ti ions that are separated from each other (Figure 6). All exhibit axial Ti-O bond lengths that increase asymmetrically by +0.04, and –0.08 Å. In addition, two equatorial Ti-O bonds become longer by +0.02 Å, and the other two equatorial bonds become shorter by -0.01 Å around the polaron. These observations are in agreement with the pioneering work of Deskins et al.,[21] who reported asymmetrically distorted bond lengths around the electron-polaron structure of anatase at U = 10 eV (VASP). Due to the shortening of the bonds in the b-lattice-vector direction (b = 3.776 Å), the lattice relaxation in the [010] direction is small, with the highest lattice relaxation observed in the c-lattice-vector direction.

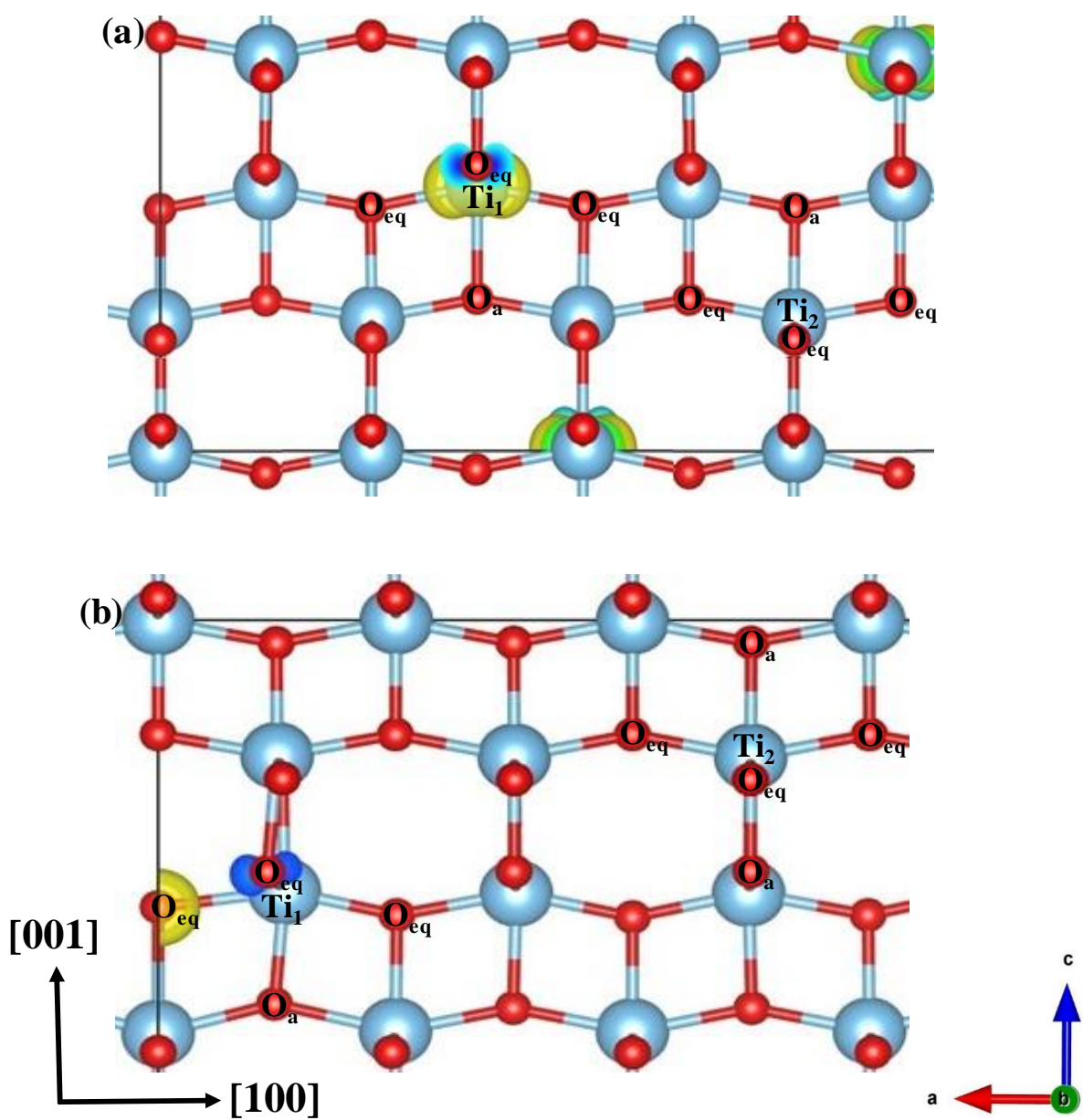

Figure 6. Hubbard-calculated anatase supercells with U = 8.0 (a): electron-polaron and (b) hole-polaron structures. Note that in (a) three electron polarons are localized on Ti ions, whereas in (b) the hole is mainly localized on a two-coordinated oxygen. Color scheme: blue, titanium; red, oxygen; yellow, polaronic wavefunction at an isosurface value of 0.005 e/Bohr$^3$. Ti-O bond distances are recorded for bonds to equatorial and axial oxygens at the polaronic site and away from it (see Table S2).

3.4.4 Hole polarons in anatase

The anatase hole polaron is localized on a two-coordinated oxygen atom; however, its wavefunction is mainly localized on a single oxygen ion. At higher U (10 eV), the hole polaron is localized on a three-coordinated oxygen site. Thus, we believe that the higher Hubbard on-site potential (U = 10 eV, VASP) was applied by Deskins et al.,[22] which resulted in a fully localized solution on three-coordinated oxygen atoms, while lower Hubbard potential values localize the hole onto two-coordinated bridging oxygens. One equatorial Ti-O bond at the hole polaron is significantly elongated (+1.00 Å), with the other diametrically opposite Ti-$O_{eq}$ bond contracts by about 0.09 Å. As a consequence, the Ti ion adjacent to the hole polaron is significantly displaced. Consequently, the highest lattice relaxation is observed in the direction of the a-lattice vector. However, the other two equatorial bonds stretch by +0.01 Å, and the smallest distortions are observed in the b-lattice-vector direction. In contrast to its electron polaron, the hole-polaron structure of anatase exhibits reduced lengths for both axial Ti-O bonds (-0.01 Å) at all polaron sites. The bond lengths stretch at the hole-polaron site (the two-coordinated oxygen atom site) by +1.00 Å and +0.05 Å.

3.5 Polaron-formation energies by hybrid DFT

Figure 7 compares the electron- and hole-polaron formation energies calculated using the PBE0 hybrid functional for brookite and anatase. Black bars represent electron-polaron formation energies, while red bars show hole-polaron formation energies. Depending on the %HF, the formation energies increase for both electron and hole polarons, as expected. Brookite and anatase both have higher hole-polaron-formation energies compared to their electron-polaron-formation energies. The energies for the formation of brookite hole polarons are ~1.6-times higher than those for the formation of electron polarons for all %HF values. Hence, there is a strong driving force for the formation of hole polarons the bulk structures of these exotic materials rather than electron polarons. An increase in the exact exchange, from 25% to 35%, results in an increase in the polaron-formation energy of approximately 0.01 eV for these brookite supercells. All of the energetics presented here are reported per polaron, as different numbers of electron and hole polarons are formed.

Although anatase produces a delocalized solution for the bulk structures studied in this work, it is worth comparing the energetics in order to get a better understanding of these systems. Here, anatase polaron-formation energies for the delocalized solution are calculated as the energy difference between $E_{deloc,non-relax}$ and $E_{deloc,relax}$. Nevertheless, the delocalized polaron energies reveal that anatase has a higher hole-polaron-forming ability, as expected from previous studies.[25, 27]

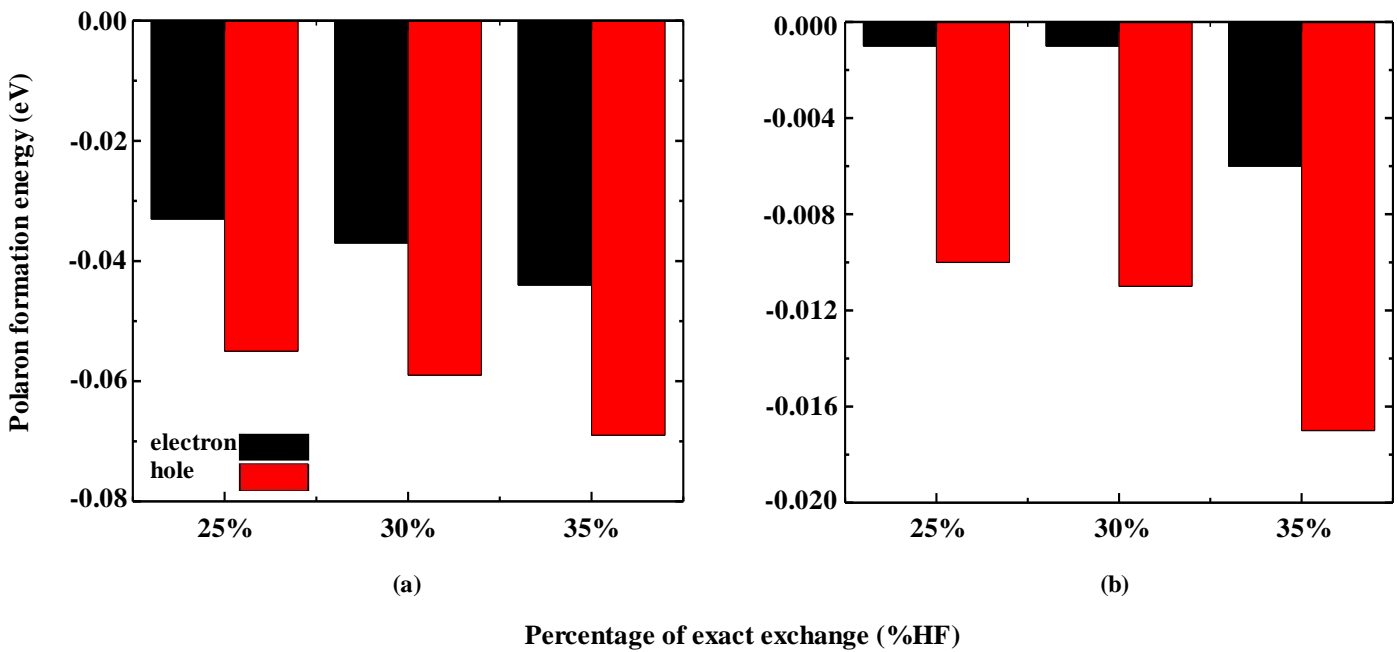

Figure 7. Polaron formation energies of (a) brookite and (b) anatase calculated by PBE0. Due to delocalization of the polarons in anatase, their energies are an order of magnitude lower than those of brookite. Note these energies are reported per polaron.

According to the literature, a 6 × 6 × 1 anatase unit cell with 432 atoms has an electron-polaron-formation energy of -0.1 eV.[29] Therefore, unit cells with 48 atoms should have very small electron-polaron-formation energies. All calculated anatase electron-polaron formation energies are less than -0.01 eV, which further validates our approach. Hole-polaron-formation energies are comparatively larger than electron-polaron-formation energies. Therefore, the driving force to produce hole polarons in anatase is also large. The hole-polaron-formation energy using 35% exact exchange is 2.8-times larger than the electron-polaron-formation energy. In addition, a recent study on rutile and anatase found that rutile exhibited maximum photocatalytic ability at a particle size of 2.5 nm, whereas 5-nm anatase particles exhibited maximum photocatalytic ability.[11] Consequently, this study further supports the observation that anatase does not produce localized electron or hole polarons that catalyze reactions at this unit-cell size. Therefore, anatase is unable to reach its optimum ability with particles of small size. In contrast, brookite produces both electron and hole polarons at this unit-cell size; hence, we can infer size-dependent photocatalytic behavior for brookite as well as anatase. Therefore, the delocalized polaronic energies in anatase shed light on its photocatalytic properties. A comparison of the electron and hole polarons of brookite with those of anatase at the high-concentration limit predicted using the hybrid functional reveals that the driving force to produce polarons in brookite is larger than that in anatase.

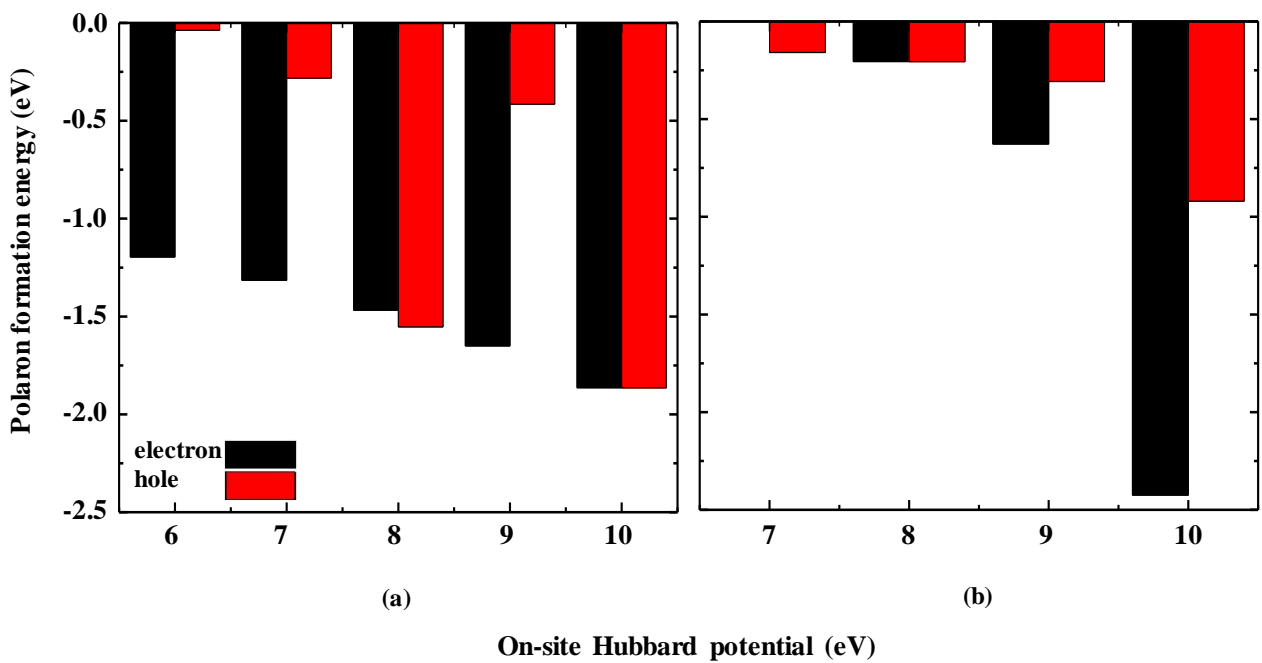

Figure 8. Polaron-formation energies of (a) brookite and (b) anatase using the on-site Hubbard model. (a) Single hole-polaron formation was observed with Hubbard U values of 8 and 10 eV, whereas in (b) the electron polarons of anatase continuously gain energy with increasing Hubbard U value. Note these energies are reported per polaron.

3.6 Polaron-formation energies by DFT + U

Brookite produces both electron and hole polarons at U = 6 eV; however, the brookite electron polaron is always localized on a single Ti ion, whereas the hole polaron is dispersed at four oxygen ions in the unit cell, except at U = 8 eV and 10 eV. Therefore, with the exception of U = 8 eV and 10 eV, the energy per polaron in a unit cell is much lower for hole polarons (see Figure 8). At these higher on-site potentials, the hole polaron tends to be mainly localized on a single oxygen ion (see Figure S2 (c)); as a result, the energy trend expected using hybrid functionals is observed. However, anatase does not form electron polarons at U = 6 and 7 eV; electron and hole polarons form at U = 8 eV, and at higher values, the electron-polaron-formation energy become much larger than the hole-polaron-formation energy. The electron and hole polarons in anatase tend to both be localized on single ions at an effective on-site Hubbard potential of 10 eV (see Figure S2 (a) and (b)). This difference is ascribable to the differences between the on-site Hubbard Coulomb potential and the Hartree-Fock functional. Hartree-Fock is an external potential in which all electrons in the system "feel", whereas the Hubbard on-site potential is applied only to a predefined set of electrons (d electrons for the titanium electron polaron, and p electrons for the oxygen hole polaron). It seems that the on-site Hubbard potential tends to affect the energetics of the d-electrons on titanium considerably more than the p electrons on oxygen. Consequently, the higher influence of the Hubbard on-site potential is reflected in the higher formation energies of the electron polarons. This influence is observed in both the brookite and anatase polymorphs. Only an on-site Hubbard potential of 8 eV reproduced the results obtained using the hybrid functional. At that value of on-site potential, brookite exhibits roughly six-times higher formation energies for both electron and hole polarons than anatase. Consequently, brookite has a higher ability to produce both types of polarons, in agreement with the hybrid-level results. Hence, the recently observed higher photocatalytic properties of the brookite polymorph are due to its superior ability to efficiently produce both types of polaron.

## 4. Conclusions

We presented electron- and hole-polaron geometries of brookite and anatase and their energetics using the hybrid PBE0 functional and on-site Hubbard potential method. The hybrid functional predicts that the electron and hole polarons formed in anatase are not localized. In contrast, brookite forms both electron and hole polarons. Despite this, the brookite electron-polaronic solution has coexisting localized and delocalized states. Hole polarons are dispersed mainly on two-coordinated oxygen ions; however, they are more pronounced and only the localized solution exists. The Hubbard on-site potential was tested for a wide range of U values (4.0–10 eV), which revealed that brookite polarons are formed at U = 6 eV and anatase polarons are formed at U = 8 eV. The brookite electron polaron was found to always be localized on a single titanium ion, whereas the hole polaron is dispersed on four oxygen atoms, as was observed in the PBE0 studies. In brookite, hole-polaron structures comparable to those predicted using the hybrid functional are observed at an on-site potential of 6 eV. The anatase electron polaron is dispersed at lower on-site potentials but becomes localized with increasing potential. The electron and hole polarons were found to be localized on single titanium and oxygen ions, respectively, at 10 eV. The hybrid-functional and DFT + U methods provided the most comparable results for anatase and brookite only at an 8 eV on-site potential. Both methods clearly predict that brookite has a higher polaron-producing driving force than anatase. This work supplements the void in the literature concerning brookite electron- and hole-polaron geometries and energetics, and provides a plausible explanation for the higher photocatalytic activity of the brookite polymorph. Furthermore, this study provides a starting point for up-scaling the system size using the more-versatile Hubbard method in order to better understand the relationship between the structures and properties of these exotic materials.

**Acknowledgements**

This study was supported by the Basic Science Research Program through the National Research Foundation of Korea (NRF) funded by the Ministry of Science, ICT and Future Planning (Nos. NRF-2017R1E1A1A01074266, NRF-2015M1A2A2057129, and NRF-2016M1A2A2937151).

**Supporting Information**

Table S1. Unperturbed, and electron- and hole-polaron structures of brookite, average Ti-O bond lengths, and deviations from the average bond lengths due to electron and hole formation calculated using PBE0 (Page S2)

Table S2. Unperturbed, and electron- and hole-polaron structures for brookite (upper) and anatase (lower), and average Ti-O bond lengths and deviations from the average bond lengths due to electron- and hole formation using on-site Hubbard U values of 6 eV for brookite and 8 eV for anatase (Page S3)

Figure S1. Fully delocalized (a) electron- and (b) hole-polaron structures of anatase. Color scheme: blue, titanium; red, oxygen; yellow, polaronic wavefunction at an isosurface value of 0.005 e/Bohr$^3$ (Page S4)

Figure S2. Disordered fully localized (a) electron- and (b) hole-polaron structures of anatase at U = 10 eV. (c) Brookite hole-polaron structure at U = 8 (and 10) eV. Color scheme: blue, titanium; red, oxygen; yellow, polaronic wavefunction at an isosurface value of 0.003 e/Bohr$^3$ (Page S4).

**2013**, *110*, 196403.

For Table of Contents use only

**Anatase PBE0**      **Brookite PBE0**

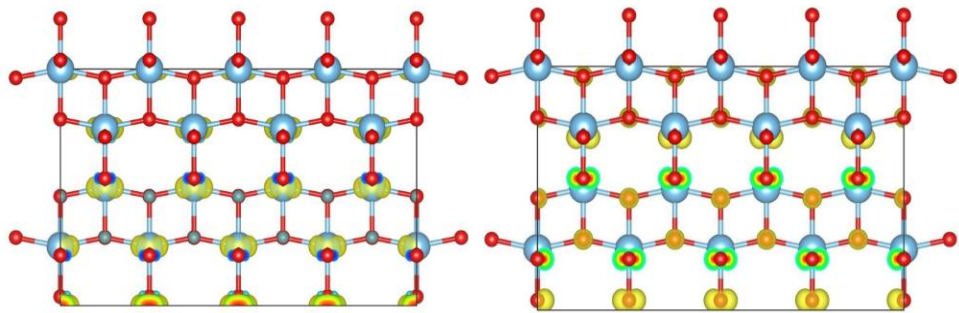
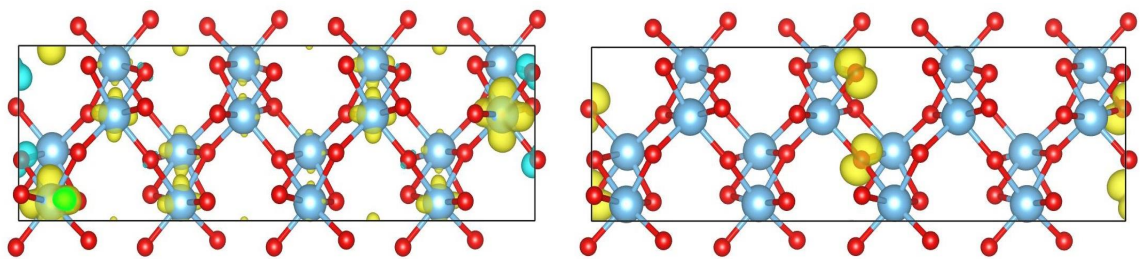

e⁻ delocalized     h⁺ delocalized     e⁻ localized-delocalized     h⁺ localized

**Anatase DFT+U**      **Brookite DFT+U**

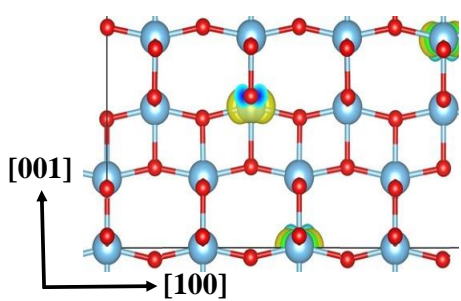
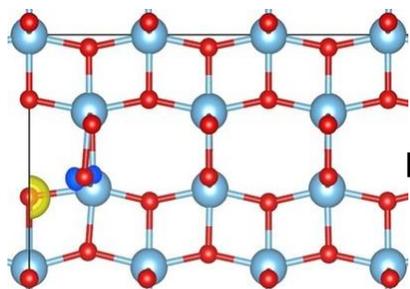
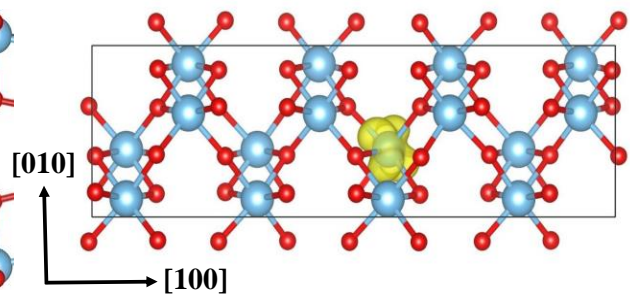
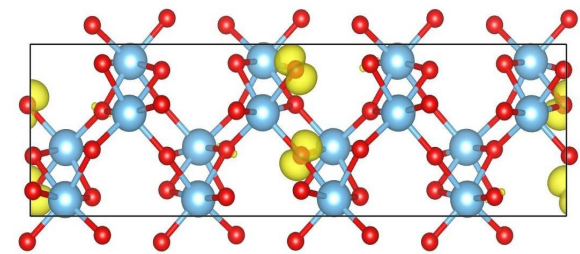

e⁻ localized     h⁺ localized     e⁻ localized     h⁺ localized

**Anatase PBE0**

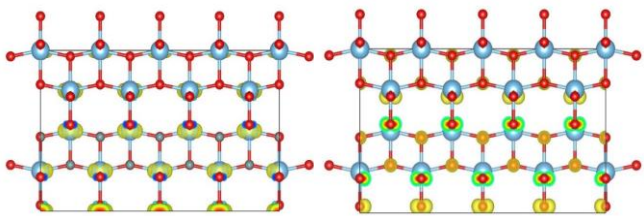

e⁻ delocalized     h⁺ delocalized

**Brookite PBE0**

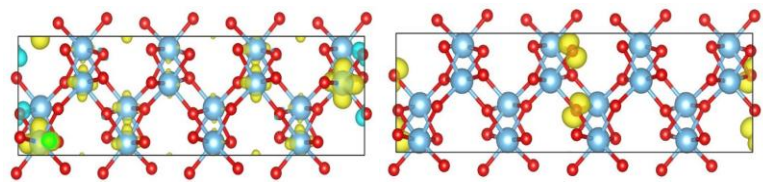

e⁻ localized-delocalized     h⁺ localized

**Anatase DFT+U**

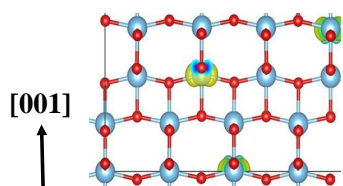 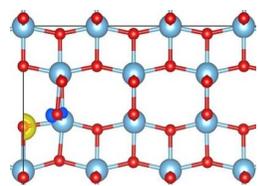

[001] ↑ → [100]

e⁻ localized     h⁺ localized

**Brookite DFT+U**

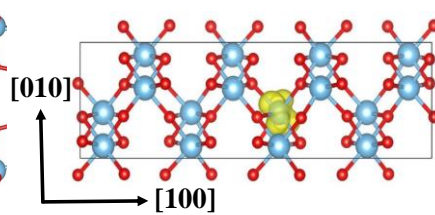 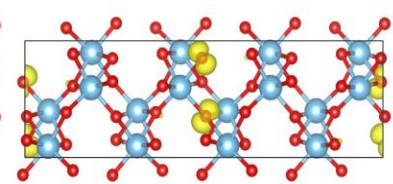

[010] ↑ → [100]

e⁻ localized     h⁺ localized